\documentclass[a4paper]{article}
\usepackage{graphicx, latexsym}
\usepackage[a4paper]{geometry}
\usepackage{natbib}
\usepackage{setspace}
\usepackage{amssymb, amsmath}
\usepackage{bm}
\usepackage{pictexwd}
\usepackage{epstopdf}
\usepackage{color}
\usepackage{colortbl}
\usepackage{array}
\usepackage{authblk}
\usepackage{mathtools}
\usepackage{bbm}
\usepackage{multirow}
\usepackage{subcaption}
\usepackage{xcolor, soul}
\usepackage{hyperref}

\sethlcolor{yellow}

\def\R{{\mathbb R}}

\def\E{{\mathbb E}}

\def\labda1{\lambda_1}
\def\labda2{\lambda_2}

\def\comment#1{\relax}

\def\=in{\mathop{\rm =}}

\renewcommand{\P}{{\mathbb P}}
\renewcommand{\E}{{\mathbb E}}

\newcommand{\Var}{{\rm Var}}

\newtheorem{thm}{Theorem}[section]

\numberwithin{equation}{section}

\begin{document}
\title{Extending the Mann-Kendall test to allow for measurement uncertainty}
\author[1,2]{Stavros Nikolakopoulos\footnote{corresponding author, email: snikolakopoulos@uoi.gr}}
\author[3]{Eric Cator}
\author[4]{Mart P. Janssen}
\affil[1]{Department of Psychology, University of Ioannina}
\affil[2]{Department of Biostatistics, University Medical Center Utrecht}
\affil[3]{Department of Mathematics, Radboud University Nijmegen}
\affil[4]{Department of Donor Medicine Research, Sanquin Research, Amsterdam}

\date{}
\maketitle

\begin{abstract}
The Mann-Kendall test for trend has gained a lot of attention in a range of disciplines, especially in the environmental sciences. One of the drawbacks of the Mann-Kendall test when applied to real data is that no distinction can be made between meaningful and non-meaningful differences in subsequent observations. We introduce the concept of partial ties, which allows inferences while accounting for (non)meaningful difference. We introduce the modified statistic that accounts for such a concept and derive its variance estimator. We also present analytical results for the behavior of the test in a class of contiguous alternatives. Simulation results which illustrate the added value of the test are presented. We apply our extended version of the test to some real data concerning blood donation in Europe.
\end{abstract}
\section{Introduction}

\subsection{Trend analysis and measurement uncertainty}\label{sec:intro}
The report "Trends and Observations on the Collection, Testing and Use of Blood and Blood Components in Europe, 2001-2011" \citep{janssen2015} presents and analyzes data provided annually to the Council of Europe by its Member States. Since 2001, data concerning blood donation in the European Union (EU) have been collected by means of a newly developed, standardized questionnaire, which was distributed to EU's Member States. Naturally, during the first years of the questionnaire administration, improvements and amendments took place, following recommendations from expert committees \citep{janssen2015}. Given this was the first time these kind of data were collected and analyzed on an annual basis, the assessment of a trend over time was important \citep{edqm,janssen2015}. As changes over time were expected (and noticed) to be extremely non-linear for some items of interest non-parametric methods were used for trend analysis.

\subsection{The Mann-Kendall test}
Let $(X,Y)$ denote a pair of random variables for which we wish to test independence and quantify ordinal association. Kendall's tau coefficient ($t$) \citep{kendall1} was originally derived to assess the degree of agreement, or correlation, between two sets of observations from such a pair of random variables. It is a non-parametric statistic, estimating population's $\tau$, also referred to as rank correlation coefficient. The $t$ statistic is a standardized version of the $S$ statistic, which in turn is calculated directly from the ranks of the observations within each of the sets of observations. 

\citet{mann}, suggested the use of $t$ for testing against the existence of a trend over time. It is a straightforward application of Kendall's $\it{\tau}$ where one of the sets of observations is timepoints, accompanied with measurements obtained at these time points. Values (and thus ranks) of the time variable are always in the natural order and $Y:=(1,2,{...},n)$ so interest lies in the quantification (and testing for) monotonic trends. The null hypothesis of the Mann-Kendall test simplifies to $(X_1,...,X_n)$ being iid. 

The Mann-Kendall test has gained a lot of attention in a range of disciplines, especially in the environmental sciences where its use has been extended in combining several sets of observations of comparable times of the year (Seasonal Kendall test for trend, \citet{hirsch}) or data obtained from different regions (Regional Kendall test for trend, \citet{regional}). The simplicity of the Mann-Kendall test makes it still popular for trend detection \citep{dawood2017,nourani2018,chen2022}.

 Methodological research in Kendall's $t$ is also an active field. \citet{perreault2022} discusses efficient computation of $t$ as well as the jackknife estimate of its variance in large datasets. \citet{nowak2021} present several contrast tests and confidence intervals that can be used to evaluate rank correlation measures in multivariate factorial designs. Non-parametric regression alternatives for modelling trends include  Local Regression, also known as LOESS or LOWESS \citep{cleveland1979}, smoothing splines \citep{hastie1987} and Gaussian Process Regression \citep{williams2006}.

 The $\it{S}$ statistic for the Mann-Kendall test is given by
\begin{equation}\label{eq:2.1}
S=\sum_{i<j}sgn(X_{j}-X_{i})
\end{equation}
where 
\[ 
sgn(x) := \left\{
\begin{array}{rr}
      -1 &{\text{   if }}x < 0,\\
      0 & {\text{   if }}x = 0,\\
      1 & {\text{   if }}x > 0,\\
\end{array} 
\right. 
\]
and $X_{.}$ refers to the variable of interest. Values of $X_{.}$ could be either the original values measured or their ranks within the sample obtained.

Kendall \citep{bible,kendall1} discusses normality of $S$, and also proves that, under $H_0$:
\begin{equation*}
\mathrm{E}(S)=0
\end{equation*}
and
\begin{equation*}
\mathrm{Var}(S)=\frac{1}{18}n(n-1)(2n+5).
\end{equation*}

Coefficient $t$ is then calculated as 

\begin{equation*}
t=\frac{2S}{n(n-1)}.
\end{equation*}

\subsection{Presence of ties}
When ties are present in the data, the formula of $Var(S)$ requires some modification. Given that there are ties of extent $w$, (e.g., if there are 3 tied groups, one duplet and 2 triplets, $w$=\{2,3,3\} ) and $w_i$ is the number of tied values per group,  the variance is calculated by \citep{bible}:
\begin{equation*}
\mathrm{Var}(S)=\frac{1}{18}\left\{ n(n-1)(2n+5)-\sum_i w_i(w_i-1)(2w_i+5) \right\}.
\end{equation*}
Behavior of $t$ and $\it{S}$ in rankings containing ties has been discussed by Kendall \citep{bible,kendall2} and others \citep{sillitto,robillard,exact}.

Based on the above, inference can be made, in the sense of statistical significance and calculating p-values, on the existence or not of a positive or negative trend over time in a given dataset.

\subsection{Introducing partially tied observations}
The biggest strength of the Mann-Kendall test is that it can provide information concerning the existence of a trend over time without any further assumptions on the character of this trend. However, one of the limitations of this approach is that $any$ increase or decrease between two observations is interpreted as such, irrespective of the size of the difference between the values compared. We found in practice (considering the blood donation data described in Section \ref{sec:intro}) that when applying the Mann-Kendall test, this lack of discrimination of the size of differences in some cases lead to a rejection of clearly visible trends. This made us explore the possibility of extending the traditional test to only value a positive or negative contribution to the test statistic in those cases where a difference in observed values is actually considered meaningful.

 From the calculation of $S$ from Equation \ref{eq:2.1}, it is obvious that pairs of values compared contribute to $S$ either +1 or -1 irrespective the magnitude of the difference between each pair's elements. One could argue though, that two observations without the exact same numerical value but with a \textit{relatively} small difference, are not reasonable to be accounted for as different and therefore should not contribute to the value of $S$. For demonstration, let us consider an example of observations on diastolic blood pressure of an individual ($DBP$) measured over 10 consecutive time points like the ones in Table \ref{T1}.

\begin{table}
\caption{Example dataset on consecutive measurements of systolic blood pressure of an individual}
\centering
\label{T1}
\begin{tabular}{l c c c c c c c c c c }
\hline \\
Time point & 1 & 2 & 3 & 4 &5 & 6 & 7 & 8 & 9 & 10  \\
\hline \\
$DBP$ &90.9&	95.2&	98.6&	95.8&	100.7&	94.9&	92.8&	101.5&	99& 98.7\\
Rank($DBP$) &1&	4&	7&	5&	9&	3&	2&	10&	8&	6\\
\hline
\\
ordered $DBP$ & 90.9&	92.8&	94.9&	95.2&	95.8&	98.6&	98.7&	99&	100.7&	101.5\\
Rank($DBP$) &1&2&3&4&5&6&7&8&9&10 \\
\hline
\end{tabular}
%\end{center}
\end{table}

As \citet{manios2007} report, diastolic blood pressure is overestimated when recorded by non-invasive methods while systolic blood pressure is underestimated when measured intravenously. One could then argue that the difference between a blood pressure of 98.6 and 98.7 is not meaningful as it is clinically irrelevant. The example could be extended to other kinds of variables with alternative sources of uncertainty (e.g. measurement error). Uncertainty could also be assumed to have one direction only, for example known rates of under or over reporting of financial or demographic measurements. To our knowledge, this problem has not been dealt with in the literature, with the only exception being a simulation study by \citet{unc}.

In light of the above, it might be worthwhile to apply a \textit{level of relevant difference (LRD)} on observations, that is, define $d$ such that a pair $(X_i,X_j)$ will be considered a tie if $|X_i-X_j|\leq d$. The value of $d$ could be defined on the basis of an expert estimate of a relevant difference, but also to a known or assumed measurement error. However, in the above example by applying a $LRD$ of $d=0.6$ we get the seemingly paradoxical situation where in the ordered sample (lower part of Table \ref{T1} and denoting by $DBP_{(i)}$ the $i$-ith ranked observation) $DBP_{(3)}=DBP_{(4)}, DBP_{(4)}=DBP_{(5)}$ but $DBP_{(3)} \neq DBP_{(5)}$. One can imagine that applying a $LRD$ rule can result in highly complex tie structures, which we will refer to as \textit{partially tied observations}. The $S$ statistic as presented in Equation \ref{eq:2.1} cannot account for such complications, and a modification for the estimation of $S$ is required.

The paper is organized as follows: In the next section, the Mann-Kendall test is extended to be able to account for partially tied observation with the key extension being the generalization of the variance formula of $S$. In section 3 a simulation study is presented to illustrate the performance of the extended test compared to the classical test. In section 4 the extended version is applied to a real data set where it is implemented within the Regional test for trend. The final section provides a discussion on the extended test.

\section{Properties of $S$ and $t$ in the presence of partially tied observations}

\subsection{Properties of $S$}
Suppose we have a sample $X_1,\ldots, X_n$ of measurements taken at subsequent times $1,2,\ldots, n$. Our test wishes to distinguish a trend in data from the null hypothesis $H_0$ that all $X_i$'s are iid. We introduce the \textit{extended} Mann-Kendall test where we allow for the above introduced $LRD$ which we denote by $d>0$ and thus we define 

\begin{equation}\label{eq:5}
S_{\rm ex} = \sum_{i<j} {\rm sign}(X_j-X_i)\mathbbm{1}_{\{|X_i-X_j|>d\}}
\end{equation}
where $ \mathbbm{1}_{\{\cdot\}}$ denotes the indicator function. The conditions of this criterion could vary as discussed above, for example the absolute could be replaced if we were interested in uncertainty towards one direction of the measurements or $>$ could be replaced by $\geq$ in Equation \ref{eq:5}, depending on the measurement units or accuracy of the measurement instrument employed.

The statistic $S_{\rm ex}$ will be strongly positive if there is a positive trend in the data, and it will be strongly negative if there is a negative trend in the data. We will base our test on a normalized version of $S_{\rm ex}$ ($t_{\rm ex}$) and for this we need to know its variance under $H_0$. 

\begin{thm}
The variance of $S_{\rm ex}$ under the null hypothesis is given by
\[ {\rm Var}(S_{\rm ex}) = (\frac13 n^3 - n^2 + \frac23 n)(\alpha_- + \alpha_+ - 2\beta) + (n^2 - n)\gamma.\]
Here,
\[ \alpha_+ = \E\left(\mathbbm{1}_{\{X_1>X_2+d\}}\mathbbm{1}_{\{X_1>X_3+d\}}\right), \ \alpha_- = \E\left(\mathbbm{1}_{\{X_1<X_2-d\}}\mathbbm{1}_{\{X_1<X_3-d\}}\right)\]
\[ \beta = \E\left(\mathbbm{1}_{\{X_1>X_2+d\}}\mathbbm{1}_{\{X_1<X_3-d\}}\right)\ \ \ \mbox{and}\ \ \ \gamma = \E\left(\mathbbm{1}_{\{X_1>X_2+d\}}\right)= \E\left(\mathbbm{1}_{\{X_1<X_2-d\}}\right).\]
\end{thm}
{\bf Proof:} It is easy to see that under the null hypothesis, 
\begin{equation*}
 \E(S_{\rm ex}|H_0) = 0.
\end{equation*}
For the variance of $S_{\rm ex}$ we therefore need to calculate 
\[\Var(S_{\rm ex}|H_0) = \E(S_{\rm ex}^2|H_0) = \sum_{i<k}\sum_{j<l}  \E\left({\rm sign}(X_k-X_i)\mathbbm{1}_{\{|X_i-X_k|>d\}}  {\rm sign}(X_l-X_j)\mathbbm{1}_{\{|X_j-X_l|>d\}}\right).\]
Define the index set 
\[ \Omega_0 = \{(i,j,k,l)\in {\mathbb N}^4 \ :\ 1\leq i<k\leq n\ \mbox{and}\ 1\leq j<l\leq n\}.\]
We split up this set into six disjoint parts:
\begin{eqnarray*}
\Omega_1 & = & \{(i,j,k,l)\in \Omega_0\ :\ i\neq j, k\neq l, i\neq l, j\neq k\}\\
\Omega_2 & = & \{(i,j,k,l)\in \Omega_0\ :\ i= j, k\neq l\}\\
\Omega_3 & = & \{(i,j,k,l)\in \Omega_0\ :\ k=l, i\neq j\}\\
\Omega_4 & = & \{(i,j,k,l)\in \Omega_0\ :\ i= l\}\\
\Omega_5 & = & \{(i,j,k,l)\in \Omega_0\ :\ j=k\}\\
\Omega_6 & = & \{(i,j,k,l)\in \Omega_0\ :\ i= j, k=l\}.\\
\end{eqnarray*}
We easily check that
\begin{eqnarray*}
\#\Omega_0 & = & \frac14 n^4 - \frac12 n^3 + \frac14 n^2\\
\#\Omega_1 & = & \frac14 n^4 - \frac32 n^3 + \frac{11}{4}n^2 - \frac32 n\\
\#\Omega_2=\#\Omega_3 & = & \frac13 n^3 - n^2 + \frac23 n\\
\#\Omega_4=\#\Omega_5 & = & \frac16 n^3 - \frac12 n^2 + \frac13 n\\
\#\Omega_6 & = & \frac12 n^2 - \frac12 n.
\end{eqnarray*}
Furthermore we define
\[ \alpha_+ = \E\left(\mathbbm{1}_{\{X_1>X_2+d\}}\mathbbm{1}_{\{X_1>X_3+d\}}\right)\ \ \ \mbox{and}\ \ \ \alpha_- = \E\left(\mathbbm{1}_{\{X_1<X_2-d\}}\mathbbm{1}_{\{X_1<X_3-d\}}\right)\]
and
\[ \beta = \E\left(\mathbbm{1}_{\{X_1>X_2+d\}}\mathbbm{1}_{\{X_1<X_3-d\}}\right)\ \ \ \mbox{and}\ \ \ \gamma = \E\left(\mathbbm{1}_{\{X_1>X_2+d\}}\right)= \E\left(\mathbbm{1}_{\{X_1<X_2-d\}}\right).\]
We then easily check the following implications under $H_0$:
\begin{eqnarray*}
(i,j,k,l)\in \Omega_1 & \Longrightarrow & \E\left({\rm sign}(X_k-X_i)1_{\{|X_i-X_k|>d\}}  {\rm sign}(X_l-X_j)1_{\{|X_j-X_l|>d\}}\right)=0\\
(i,j,k,l)\in \Omega_2\cup\Omega_3 & \Longrightarrow & \E\left({\rm sign}(X_k-X_i)1_{\{|X_i-X_k|>d\}}  {\rm sign}(X_l-X_j)1_{\{|X_j-X_l|>d\}}\right)=\alpha_-+\alpha_+-2\beta\\
(i,j,k,l)\in \Omega_4\cup\Omega_5 & \Longrightarrow & \E\left({\rm sign}(X_k-X_i)1_{\{|X_i-X_k|>d\}}  {\rm sign}(X_l-X_j)1_{\{|X_j-X_l|>d\}}\right)=2\beta - \alpha_- - \alpha_+\\
(i,j,k,l)\in \Omega_6 & \Longrightarrow & \E\left({\rm sign}(X_k-X_i)1_{\{|X_i-X_k|>d\}}  {\rm sign}(X_l-X_j)1_{\{|X_j-X_l|>d\}}\right)=2\gamma\\
\end{eqnarray*}
Our conclusion is that under $H_0$, 
\[ \Var(S_{\rm ex}|H_0) = (\frac13 n^3 - n^2 + \frac23 n)(\alpha_- + \alpha_+ - 2\beta) + (n^2 - n)\gamma. \ \ \ \ \Box\]

Note that $\mathbbm{1}_{\{X_1>X_2+d\}}$ and $\mathbbm{1}_{\{X_1>X_3+d\}}$ are positively correlated, $\mathbbm{1}_{\{X_1<X_2-d\}}$ and $\mathbbm{1}_{\{X_1<X_3-d\}}$ are also positively correlated, but $\mathbbm{1}_{\{X_1>X_2+d\}}$ and $\mathbbm{1}_{\{X_1>X_3-d\}}$ are negatively correlated (of course, this also holds if we use any triplet $X_i, X_j, X_k$, for different values of $i,j$ and $k$). This implies
\[ \min(\alpha_-,\alpha_+)\geq \gamma^2\geq \beta.\]
Also, if $d=0$ and the $X_i$'s are continuous random variables, we see that $\alpha_-=\alpha_+=\frac13$, $\beta=\frac16$ and $\gamma=\frac12$. This leads to the classical variance formula for $S$ without ties.

\subsection{Estimating the variance of $S_{\rm ex}$}

A natural way to estimate $\Var(S_{\rm ex}|H_0)$ would be to use unbiased estimates of $\alpha_-,\alpha_+, \beta $ and $\gamma$. Since these constants are defined as expectations, we will estimate them by averages. Also, since under $H_0$ all variables are iid, we will use a symmetric average:
\begin{eqnarray*}
\hat{\alpha}_+ & = &\frac{1}{n(n-1)(n-2)} \sum_{\{i,j,k\ :\ i\neq j\neq k\}} \left(\mathbbm{1}_{\{X_i>X_j+d\}}\mathbbm{1}_{\{X_i>X_k+d\}}\right)\\
 & = &\frac{1}{n(n-1)(n-2)}\sum_{i=1}^n \sum_{j\neq k} \left(\mathbbm{1}_{\{X_i>X_j+d\}}\mathbbm{1}_{\{X_i>X_k+d\}}\right)\\
 & =& \frac{1}{n(n-1)(n-2)}\sum_{i=1}^n \left(\sum_{j=1}^n \mathbbm{1}_{\{X_i>X_j+d\}}\cdot \sum_{k=1}^n \mathbbm{1}_{\{X_i>X_k+d\}} - \sum_{j=1}^n \mathbbm{1}_{\{X_i>X_j+d\}}\right)\\
 & = & \frac{1}{n(n-1)(n-2)}\sum_{i=1}^n \left(u_i^2-u_i \right).
\end{eqnarray*}
Here we define
\[ u_i = \#\{j\in\{1,\ldots, n\}\ :\ X_i>X_j+d\}\ \ \ \mbox{and}\ \ \ v_i = \#\{j\in\{1,\ldots, n\}\ :\ X_i<X_j-d\}.\]
Similar calculations show
\begin{eqnarray*}
\hat{\alpha}_- & = & \frac{1}{n(n-1)(n-2)} \sum_{i=1}^n \left(v_i^2-v_i\right)\\
\hat{\beta} & = & \frac{1}{n(n-1)(n-2)}\sum_{i=1}^n u_iv_i\\
\hat{\gamma} & = & \frac{1}{n(n-1)}\sum_{i=1}^n u_i = \frac{1}{n(n-1)}\sum_{i=1}^n v_i.
\end{eqnarray*}
Finally we get an estimator for the variance of $S_{\rm ex}|H_0$:
\[ \widehat\Var(S_{\rm ex}|H_0) = \frac13\sum_{i=1}^n (u_i-v_i)^2 + \frac13 \sum_{i=1}^n u_i.\]
Our adapted Mann-Kendall test statistic $T$ therefore becomes
\begin{equation}\label{eq:T}
 T  = \frac{S_{\rm ex}}{\sqrt{ \widehat\Var(S_{\rm ex})}} = \frac{\sum_{i<j} {\rm sign}(X_j-X_i)\mathbbm{1}_{\{|X_i-X_j|>d\}}}{\sqrt{ \frac13\sum_{i=1}^n (u_i-v_i)^2 + \frac13 \sum_{i=1}^n u_i}}.
 \end{equation}

\subsection{Contiguous alternatives}

In Section \ref{Choice of value for $d$} we will discuss some guidelines in choosing $d$. In this section we wish to study the power of our test for different choices of $d$ in a more theoretical manner. A good way to do this, is by using what is known as {\it contiguous alternatives} to the null hypothesis that there is no trend, by which we mean the following: suppose $Z_1,\ldots , Z_n$ are iid random variables, choose $\lambda \in \R$ and define 
\[ X_i = \lambda\,\frac{i}{n^{3/2}} + Z_i.\]
This means that the $X_i$'s have a trend (and so we are not in the null hypothesis), but as the sample size $n$ grows, the trend becomes smaller: in the end-point the difference decreases like $n^{-1/2}$, precisely enough to still be noticeable by the classic Mann-Kendall test. The advantage of such alternatives is that we are able to write down the asymptotic power of our test in a more or less closed form, so that we can compare its power for different choices of $d$ in our test. We need some notation: suppose $f(\cdot)$ is the density of $Z_1$, and define $g(\cdot)$ as the density of $Z_1-Z_2$. This means that
\[ g(z) = \int_{\R} f(x+z)f(x)\,dx.\]
For simplicity we will assume that $g(0)<+\infty$, although this is not strictly necessary. It follows that $g$ is symmetric (so $g(-z)=g(z)$) and $g$ has a maximum at $z=0$, simply because Cauchy-Schwartz shows that
\[ g(z) \leq \sqrt{\int_{\R} f(x)^2\,dx\cdot \int_{\R} f(x+z)^2\,dx} = g(0).\]
We can now show the following result on the asymptotic distribution of our test statistic $T$:
\begin{thm}
Given the notation above, the limiting distribution of $T$ as $n\to \infty$ is given by
\[ T \sim N\left(\frac{\lambda g(d)}{\sqrt{3(\alpha_-+\alpha_+-2\beta)}}, 1\right).\]
\end{thm}
{\bf Proof:} We introduce the subscript $X$ or $Z$, meaning that we are either using the sample $X_1,\ldots,X_n$, or the coupled sample $Z_1,\ldots,Z_n$ which satisfies the null hypothesis (so $T=T_X$). We will start by showing that asymptotically, $\widehat{\rm Var}(S_{{\rm ex},X})/ \widehat{\rm Var}(S_{{\rm ex},Z}) \longrightarrow 1$.
Note that
\begin{eqnarray*}
|\mathbbm{1}_{\{X_i>X_j+d\}}\mathbbm{1}_{\{X_i>X_k+d\}}-\mathbbm{1}_{\{Z_i>Z_j+d\}}\mathbbm{1}_{\{Z_i>Z_k+d\}}| & \leq & \mathbbm{1}_{\{Z_i-Z_j\in [d-|\lambda (j-i)|n^{-3/2}, d+|\lambda (j-i)|n^{-3/2}]\}}\ +\\
& & \ \ \  \mathbbm{1}_{\{Z_i-Z_k\in [d-|\lambda (k-i)|n^{-3/2}, d+|\lambda (k-i)|n^{-3/2}]\}} \\
& \leq &\mathbbm{1}_{\{Z_i-Z_j\in [d-|\lambda|n^{-1/2}, d+|\lambda|n^{-1/2}]\}} + \\
& & \ \ \  \mathbbm{1}_{\{Z_i-Z_k\in [d-|\lambda|n^{-1/2}, d+|\lambda|n^{-1/2}]\}}.
\end{eqnarray*}
The expected value of the right-hand side is $O(n^{-1/2})$ (since this is the width of the intervals and $g$, the density of $Z_i-Z_j$, is bounded by $g(0)$), and since $\hat{\alpha}_{+,X} -  \hat{\alpha}_{-,Z}$ is bounded by the average of these bounded random variables, we conclude that
\begin{equation}\label{eq:varX=Z}
\hat{\alpha}_{+,X} =  \hat{\alpha}_{-,Z} + O_p(n^{-1/2}).
\end{equation}
The same conclusion holds for $\hat{\alpha}_-$ and $\hat{\beta}$, which means that
\[ \frac{\widehat{\Var}(S_{{\rm ex},X})}{ (\frac13 n^3 - n^2 + \frac23 n)} \to \alpha_- + \alpha_+ - 2\beta\ \ \ \ (\mbox{as }n\to \infty),\]
where the constants are defined for the distribution of $Z_1$. In other words, the contiguous alternative does not significantly alter the estimation of the variance of $S_{\rm ex}$. However, the alternative does significantly change the expected value of $S_{\rm ex}$. Under the null hypothesis, we know that $S_Z = O_p(n^{3/2})$. Now let us calculate $\E(S_X)$:
\begin{eqnarray*}
\E(S_X) & = & \sum_{i<j} \P(X_i<X_j-d) - \P(X_i>X_j+d)\\
& = &  \sum_{i<j} \P(Z_i-Z_j<-d + \lambda (j-i)n^{-3/2}) - \P(Z_i-Z_j>d +  \lambda (j-i)n^{-3/2})\\
& = &  \sum_{i<j} \P(Z_i-Z_j\in [d -  \lambda (j-i)n^{-3/2}, d +  \lambda (j-i)n^{-3/2}])\\
& = & (\sum_{i<j} 2g(d)\lambda(j-i)n^{-3/2})(1+o(1))\\
& = & \frac13 \lambda g(d)n^{-3/2}(1+o(1)).
\end{eqnarray*}
Here we used $Z_i-Z_j$ has the same distribution as $Z_j-Z_i$. As you can see, the difference is of the order square root of the variance. Furthermore, from \eqref{eq:varX=Z} we conclude that
\[ T_X \sim N\left(\frac{\lambda g(d)}{\sqrt{3(\alpha_-+\alpha_+-2\beta)}}, 1\right).\ \ \ \Box\]

To investigate the power of our test against this contiguous alternative as a function of $d$ (remember, we reject the null hypothesis if $|T_X|\geq z_{\alpha/2}$, the normal critical value), it is therefore sufficient to investigate $\E_d(T_X)$, the expected value of $T_X$ as a function of $d$, given by
\[ \E_d(T_X) = \frac{\lambda g(d)}{\sqrt{3(\alpha_-+\alpha_+-2\beta)}}.\]
Here, we have left out the lower order terms as $n$ goes to infinity. The power of our test is clearly an increasing function of $\E_d(T_X)$: if $\E_d(T_X)=0$, the power equals $\alpha$, and if $\E_d(T_X)\to \infty$, then the power increases to $1$. We start by investigating the behaviour around $d=0$. It is clear that $\E_d(T_X)$ is a symmetric function of $d$. After some tedious calculus, it turns out that this function has a positive second derivative at $0$, if and only if
\begin{equation}\label{eq:possecder}
\int f(z)^2\,dz\cdot \int f(z)^3\,dz > \frac16 \int f'(z)^2\,dz.
\end{equation}
where $f'(z)$ denotes the first derivative of $f(z)$. If this is the case, it means that we can gain power even asymptotically by choosing $d$ slightly larger than $0$. There exist distributions $f$ of $Z_i$ that do not satisfy Condition \eqref{eq:possecder}, but we will see that it is satisfied if the $Z_i$'s are normally distributed.
Considering large values of $d$ is not a good idea, since the number of "ties" will explode.  A notable exception is when $X_1$ has a uniform distribution on, for example, $[0,1]$. In that case, it would be good to choose $d$ slightly larger than 1, and the power with respect to the contiguous alternative would go to 1. This is of course due to the fact that if we would see a difference between two $X_i$'s larger than 1, we could always reject the null hypothesis. 

As mentioned before, an interesting case is if all $Z_i$'s have a normal distribution with standard deviation $\sigma$. Figure \ref{fig:ET} shows $\E_d(T_X)$ as a function of $d$ for $\sigma=1$ and $\lambda=1$. 

\begin{figure}[h]
\[
\beginpicture
\footnotesize
\setcoordinatesystem units <0.3\textwidth,4\textwidth>
  \setplotarea x from 0 to 3, y from 0.23 to 0.285
  \axis bottom
    shiftedto y=0.23
    ticks numbered from 0 to 3 by 0.5
    /
  \axis left
    shiftedto x=0.0
    ticks numbered from 0.232 to 0.284 by 0.013
    /
\setlinear
\plot
0	0.282094791773878
0.01	0.282095515732266
0.02	0.282097687339649
0.03	0.2821013057927
0.04	0.282106369752601
0.05	0.282112877345125
0.06	0.282120826160766
0.07	0.282130213254888
0.08	0.282141035147924
0.09	0.282153287825603
0.1	0.282166966739213
0.11	0.282182066805907
0.12	0.282198582409038
0.13	0.282216507398541
0.14	0.282235835091342
0.15	0.282256558271815
0.16	0.282278669192279
0.17	0.282302159573529
0.18	0.282327020605412
0.19	0.282353242947448
0.2	0.282380816729492
0.21	0.282409731552434
0.22	0.28243997648896
0.23	0.282471540084343
0.24	0.282504410357288
0.25	0.282538574800833
0.26	0.282574020383287
0.27	0.282610733549226
0.28	0.282648700220547
0.29	0.282687905797564
0.3	0.282728335160172
0.31	0.282769972669056
0.32	0.28281280216697
0.33	0.282856806980069
0.34	0.282901969919303
0.35	0.282948273281879
0.36	0.282995698852781
0.37	0.283044227906362
0.38	0.283093841208005
0.39	0.283144519015848
0.4	0.283196241082586
0.41	0.283248986657348
0.42	0.283302734487643
0.43	0.28335746282139
0.44	0.283413149409023
0.45	0.283469771505681
0.46	0.283527305873474
0.47	0.283585728783844
0.48	0.283645016020002
0.49	0.283705142879463
0.5	0.283766084176666
0.51	0.283827814245688
0.52	0.283890306943052
0.53	0.283953535650634
0.54	0.284017473278664
0.55	0.284082092268831
0.56	0.28414736459749
0.57	0.284213261778967
0.58	0.284279754868981
0.59	0.284346814468162
0.6	0.284414410725686
0.61	0.284482513343021
0.62	0.28455109157778
0.63	0.284620114247699
0.64	0.284689549734719
0.65	0.284759365989199
0.66	0.28482953053424
0.67	0.284900010470134
0.68	0.284970772478937
0.69	0.285041782829161
0.7	0.285113007380601
0.71	0.285184411589281
0.72	0.285255960512533
0.73	0.285327618814204
0.74	0.285399350769998
0.75	0.285471120272945
0.76	0.285542890839006
0.77	0.285614625612812
0.78	0.285686287373539
0.79	0.285757838540922
0.8	0.285829241181395
0.81	0.285900457014383
0.82	0.285971447418722
0.83	0.286042173439222
0.84	0.286112595793368
0.85	0.286182674878158
0.86	0.286252370777088
0.87	0.286321643267265
0.88	0.286390451826671
0.89	0.286458755641564
0.9	0.286526513614006
0.91	0.286593684369553
0.92	0.286660226265061
0.93	0.286726097396646
0.94	0.286791255607772
0.95	0.286855658497479
0.96	0.286919263428747
0.97	0.286982027536997
0.98	0.287043907738715
0.99	0.287104860740227
1	0.287164843046582
1.01	0.287223810970584
1.02	0.287281720641939
1.03	0.287338528016534
1.04	0.287394188885839
1.05	0.287448658886426
1.06	0.287501893509618
1.07	0.287553848111241
1.08	0.287604477921504
1.09	0.287653738054982
1.1	0.287701583520717
1.11	0.287747969232415
1.12	0.287792850018756
1.13	0.287836180633801
1.14	0.287877915767496
1.15	0.287918010056267
1.16	0.287956418093716
1.17	0.287993094441387
1.18	0.288027993639633
1.19	0.288061070218545
1.2	0.288092278708976
1.21	0.288121573653612
1.22	0.288148909618131
1.23	0.288174241202417
1.24	0.288197523051822
1.25	0.288218709868501
1.26	0.288237756422779
1.27	0.28825461756457
1.28	0.288269248234834
1.29	0.288281603477069
1.3	0.288291638448826
1.31	0.28829930843326
1.32	0.288304568850687
1.33	0.288307375270165
1.34	0.288307683421073
1.35	0.288305449204705
1.36	0.288300628705849
1.37	0.288293178204368
1.38	0.288283054186758
1.39	0.288270213357694
1.4	0.288254612651547
1.41	0.28823620924387
1.42	0.28821496056285
1.43	0.288190824300713
1.44	0.288163758425086
1.45	0.288133721190299
1.46	0.288100671148633
1.47	0.288064567161499
1.48	0.288025368410543
1.49	0.287983034408682
1.5	0.287937525011044
1.51	0.287888800425834
1.52	0.287836821225089
1.53	0.287781548355349
1.54	0.287722943148209
1.55	0.287660967330766
1.56	0.287595583035943
1.57	0.287526752812701
1.58	0.287454439636108
1.59	0.287378606917288
1.6	0.287299218513219
1.61	0.287216238736398
1.62	0.287129632364348
1.63	0.287039364648976
1.64	0.286945401325766
1.65	0.286847708622817
1.66	0.286746253269701
1.67	0.286641002506157
1.68	0.2865319240906
1.69	0.286418986308453
1.7	0.286302157980291
1.71	0.28618140846979
1.72	0.286056707691489
1.73	0.285928026118345
1.74	0.285795334789096
1.75	0.285658605315408
1.76	0.28551780988882
1.77	0.285372921287476
1.78	0.285223912882637
1.79	0.285070758644983
1.8	0.284913433150686
1.81	0.284751911587265
1.82	0.284586169759213
1.83	0.284416184093401
1.84	0.284241931644239
1.85	0.284063390098627
1.86	0.283880537780646
1.87	0.283693353656042
1.88	0.283501817336448
1.89	0.283305909083388
1.9	0.283105609812036
1.91	0.282900901094731
1.92	0.282691765164259
1.93	0.282478184916896
1.94	0.282260143915206
1.95	0.282037626390599
1.96	0.281810617245656
1.97	0.281579102056204
1.98	0.281343067073162
1.99	0.281102499224141
2	0.280857386114808
2.01	0.280607716030018
2.02	0.280353477934705
2.03	0.28009466147454
2.04	0.279831256976358
2.05	0.279563255448355
2.06	0.279290648580051
2.07	0.279013428742033
2.08	0.278731588985463
2.09	0.278445123041377
2.1	0.27815402531975
2.11	0.277858290908349
2.12	0.277557915571376
2.13	0.277252895747887
2.14	0.276943228550012
2.15	0.276628911760963
2.16	0.27630994383284
2.17	0.275986323884239
2.18	0.27565805169766
2.19	0.275325127716722
2.2	0.274987553043193
2.21	0.274645329433828
2.22	0.274298459297029
2.23	0.273946945689324
2.24	0.273590792311674
2.25	0.273230003505605
2.26	0.272864584249179
2.27	0.2724945401528
2.28	0.272119877454858
2.29	0.271740603017226
2.3	0.2713567243206
2.31	0.2709682494597
2.32	0.270575187138323
2.33	0.270177546664265
2.34	0.269775337944102
2.35	0.269368571477855
2.36	0.268957258353519
2.37	0.268541410241477
2.38	0.268121039388805
2.39	0.267696158613452
2.4	0.267266781298329
2.41	0.266832921385283
2.42	0.266394593368986
2.43	0.265951812290716
2.44	0.265504593732065
2.45	0.265052953808548
2.46	0.264596909163141
2.47	0.264136476959739
2.48	0.263671674876544
2.49	0.263202521099384
2.5	0.262729034314964
2.51	0.26225123370407
2.52	0.261769138934703
2.53	0.261282770155173
2.54	0.260792147987134
2.55	0.260297293518591
2.56	0.259798228296848
2.57	0.259294974321436
2.58	0.258787554037004
2.59	0.258275990326173
2.6	0.257760306502375
2.61	0.257240526302671
2.62	0.256716673880536
2.63	0.256188773798651
2.64	0.255656851021662
2.65	0.255120930908948
2.66	0.254581039207369
2.67	0.254037202044024
2.68	0.253489445919002
2.69	0.252937797698136
2.7	0.252382284605773
2.71	0.251822934217539
2.72	0.251259774453128
2.73	0.250692833569095
2.74	0.25012214015168
2.75	0.249547723109632
2.76	0.248969611667072
2.77	0.248387835356371
2.78	0.247802424011055
2.79	0.24721340775874
2.8	0.246620817014095
2.81	0.246024682471841
2.82	0.245425035099777
2.83	0.244821906131848
2.84	0.244215327061253
2.85	0.243605329633577
2.86	0.242991945839984
2.87	0.242375207910439
2.88	0.241755148306975
2.89	0.241131799717011
2.9	0.240505195046707
2.91	0.239875367414378
2.92	0.239242350143943
2.93	0.238606176758439
2.94	0.237966880973569
2.95	0.237324496691318
2.96	0.236679057993607
2.97	0.236030599136014
2.98	0.235379154541536
2.99	0.234724758794416
3	0.234067446634022
/
\endpicture
\]
\caption{$\E_d(T_X)$ (vertical axis) against $d$.}\label{fig:ET}
\end{figure}

The expected value for $T$ is maximal around $d=1.34\sigma$ (as $g(d)$ and thus the mean of the distribution of the test statistic is invariant to $\sigma$ for $d=c \sigma$ and only depends on c), but there is only a very slight increase in power. Note that if you pick $d$ too large, the power quickly drops, and furthermore, the increase of power for small $d$ is only quadratic in $d$ (this holds for any differentiable $f$, and if Condition \eqref{eq:possecder} is violated, there would be a quadratic \emph{decrease} of power). In Section \ref{Choice of value for $d$} we will discuss some guidelines in the choice of $d$ for real data sets.

\subsection{Properties of $t$}

For the standard $S$ statistic (no $LRD$ applied) and when no ties are present in the data, then $max(S)=\frac{1}{2}n(n-1)$ and $t$ is defined as :
\begin{equation*}
t=\frac{2S}{n(n-1)}.
\end{equation*}

In  case of tied ranks, $t_{a}$ or $t_{b}$ can be used. If by $c_{ij}$ we denote the score allotted to any pair of measurements subject to the conditions that $c_{ij}$=-$c_{ji}$ and $c_{ij}$=0 if $i=j$ (so that, for $S$ and given the natural order ($i<j$), $c_{ij}$=1). For the variable representing time, we denote the score $e_{ij}$, subject to the same conditions. Then, $t_{a}$ and $t_{b}$ are defined as follows: $t_{a}$ is still $S$ divided by $\frac{1}{2}n(n-1)$ while in $t_{b}$ the denominator is replaced by $\frac{1}{2}\sqrt{\sum {c^2}_{ij} \sum {e^2}_{ij}}$. For our extended Mann-Kendall test, it is straightforward to see that $\sum {c^2}_{ij}=\sum_iu_i+\sum_iv_i=2\sum_iu_i=2\sum_iv_i$. It then follows  that in the case of the Mann-Kendall test where $\sum{e^2}_{ij}=n(n-1)$ since $e_{ij}=1$ ,$\forall i \neq j$ for the time variable rankings, $t_{b}$ will take the form:

\begin{eqnarray*}
t_{b_{\rm ex}}&=&\frac{S_{\rm ex}}{\frac{1}{2}\sqrt{\left\{2\sum_iu_i\right\}\left\{n(n-1)\right\}}}.
\end{eqnarray*}

We refer the interested reader to the discussion provided by Kendall and Dickinson (1990, Chapter 3) on whether to use $t_{a}$ or $t_{b}$.

\subsection{Distribution of $S$ and $T$}\label{sec:dis}
\citet{bible} note that "the distribution of $T$ for any fixed number of ties tends to normality as $n$ increases, and there is probably little important error involved in using the normal approximation for $n\geq10$, unless the ties are very extensive or very numerous in which case a special investigation may be necessary". This conjecture was verified by \citet{burr} who examined tied ranks in both rankings and the normal approximation was found not to be adequate only in cases where only a few of the extreme values of $S$ are significant, e.g. the distribution of $S$ is sparsely represented in the critical region. It is suggested that these concerns should be taken into consideration when applying a $LRD$ to a sample and the structure of ties should be monitored.

The distribution of $S_{\rm ex}$ (and $T_{\rm ex}$ since it is a multiple of $S_{\rm ex}$) is always symmetric and centered around 0 (under $H_0$) when we are dealing with partially tied observations, and the ties only impact its variance. Therefore, the continuity correction suggested by \citet{bible} is still applicable. This results in the $Z$ statistic to be used for testing against trend where:

\begin{equation}\label{eq:Z}
Z= \left\{
\begin{array}{rl}
\frac{S_{\rm ex}+1}{\sigma_{S_{\rm ex}}}& \text{if } S_{\rm ex}< 0\\
\\
0 & \text{if } S_{\rm ex} = 0\\
\\
\frac{S_{\rm ex}-1}{\sigma_{S_{\rm ex}}} & \text{if } S_{\rm ex} > 0
\end{array} \right.
\end{equation}

and $\sigma_{S_{\rm ex}}=\sqrt{\mathrm{Var}(S_{\rm ex})}$ so that $Z|H_0 \sim N(0,1)$.

\section{Simulation study}
We conducted two simulation studies. First we studied the operational characteristics of the extedned test under linear or quadratic terms and variable values of $d$. Second, we assessed the minimum Effective Sample Size required for the normality assumption to hold. We report on them below. 

\subsection{Operational characteristics}
A simulation study was conducted to assess the operational characteristics of the proposed extended version of the Mann-Kendall test and its added value compared to the standard test. The study aims to evaluate the performance of the test in the settings of linear and quadratic trends over time. The general data-generating model is:

\begin{align*}
Y=\theta X^p + \epsilon
\end{align*}
where $Y$ denotes the variable of interest, $X:=\{1,\dots,n\}$ is the (fixed) time measurements (corresponding to the ranks of the time variable) and $\epsilon$ is the  measurement error associated with this process. The simulation parameters include $\theta \in \{0,1\}$ ,$p\in\{1,2\}$, $n\in\{20,30\}$ and $\epsilon$ is generated from either Normal or Uniform distributions with mean 0 and standard deviations $\sigma^{1/p}\in \{10,15,20\}$. The parameters of the Uniform distributions $U(a,b)$ are thus calculated as $a=-\sigma\sqrt{3}$ and $b=-a$. This corresponds to $\sigma$ ranging from 10 to 400 (for $p=2$ and $\sigma^{1/p}=20$) and $a$ ranging from -17 to -690. For each scenario, the extended Mann-Kendall test is applied, with $LRD$'s $d/\sigma \in \{0,0.5,1,1.5,2\}$. Thus, $d$ is supposed to be determined relative to the measurement error, where $d$=0 refers to the standard Mann-Kendall test. Testing is conducted at the $5\%$ significance level, two-sided. The choice of the sample sizes and effect sizes was based on the pursuit to explore the validity of the normality assumption of the extended test in borderline small sample sizes in combination with the concept of partial ties. All simulations were conducted with the R language for statistical computing. 

The simulations results are presented in Tables \ref{tab:Res} and \ref{tab:Resb}. Table \ref{tab:Res} presents empirical power values (and thus type I error for $\theta=0$) for the explored scenarios while Table \ref{tab:Resb} presents the average proportion of ties observed in each scenario, $\pi_{t}=\E\left(\frac{maxS-\sum_i{u_i}}{maxS}\right)$, where $\max S=\frac{1}{2}n(n-1)$. In order to evaluate the performance of the extended test we refer to each respective cell from the two tables simultaneously, so that validity of the normality assumption is assessed together with the resulting operational characteristics.

By looking at Tables \ref{tab:Res1} and \ref{tab:Res2} jointly, the percentiles of the null distribution of our extended test statistics seem to align with the ones of the assumed normal distribution (type I error rates close to nominal), for moderate proportions of ties observed. Thus, normality assumption seems to hold for $\pi_{t}<0.6$. The approximation might be judged adequate for even larger proportions of ties when the error distribution is assumed to be Uniform. The latter however seems to be affected by both the sample size and the extent of tied values, therefore attention should be paid, as to the standard version of the test. The extended version of the Mann-Kendall test is doing exactly what is intended, namely increasing the power of the Mann-Kendall test in the presence of measurement error. For normally distributed errors and a sample size of 30, (Table \ref{tab:Res1}) applying a $LRD$ increases power for values of $d/\sigma$ up to 1.5 and a slight drop is observed for $d/\sigma=2$. The drop could however be attributed to the inadequacy of the normal approximation due to the large number of ties when applying such a (relatively) large $d$, as the type I error is shown to deviate from the nominal level as well.  

The gains in power seem to be even more substantial in the case of uniform distributed errors (Table \ref{tab:Res2}) in combinations of sample sizes and $\pi_{t}$'s where the normal approximation seems to hold (type I error levels close to nominal). In such cases there is a substantial increase in power for large measurement error, when this is taken into account through the application of a $LRD$. It is evident from the simulation results that both the magnitude and the distribution of the measurement error should be taken into account so that an optimal $d$ is chosen for the analysis. This point is elaborated on in more detail in the following section. 

\begin{table}[]
\caption{Probability of rejection per simulation scenario. Data are generated from $\theta X^p + \epsilon$ where $\epsilon$ follows the respective distribution. $d / \sigma$=0 corresponds to the standard Mann-Kendall test with no $LRD$.  See text for details.  }\label{tab:Res}
\begin{subtable}[t]{.45\linewidth}
\centering
\caption{Normal distribution}\label{tab:Res1}
\begin{tabular}[t]{llrccc}
&&&\multicolumn{3}{c}{$\theta$}\\
&&&0&\multicolumn{2}{c}{1}\\
\cline{4-6}
&&&&\multicolumn{2}{c}{$p$}\\
\cline{5-6}
$n$&$\sigma^{\frac{1}{p}}$&$d / \sigma$&&1&2\\
\hline
\multirow{15}{*}{20}&\multirow{5}{*}{10}&0.0&	0.047&	0.632&	0.996 \\
&&0.5&	0.044&	0.643&	0.997 \\
&&1.0&	0.042&	0.643&	0.997 \\
&&1.5&	0.038&	0.619&	0.997 \\
&&2.0&	0.027&	0.553&	0.996 \\
\cline{2-6}
&\multirow{5}{*}{15}&0.0	&	0.047	&0.333&	0.564\\
&&0.5&	0.044&	0.337&	0.574\\
&&1.0&	0.042&	0.334&	0.573\\
&&1.5&	0.038&	0.313&	0.552\\
&&2.0&	0.027&	0.259&	0.488\\
\cline{2-6}
&\multirow{5}{*}{20}&0.0&	0.047&	0.207&	0.222\\
&&0.5&	0.045&	0.208&	0.223\\
&&1.0&	0.043&	0.205&	0.220\\
&&1.5&	0.038&	0.189&	0.204\\
&&2.0&	0.027&	0.151&	0.164\\
\hline
\multirow{15}{*}{30}&\multirow{5}{*}{10}&0.0&		0.049	&0.992	&1.000\\
&&0.5&	0.047&	0.993&	1.000\\
&&1.0&	0.046&	0.994&	1.000\\
&&1.5&	0.044&	0.993&	1.000\\
&&2.0&	0.038&	0.990&	1.000\\
\cline{2-6}
&\multirow{5}{*}{15}&0.0&		0.049&	0.833&	1.000\\
&&0.5&	0.047&	0.840&	1.000\\
&&1.0&	0.046&	0.843&	1.000\\
&&1.5&	0.044&	0.835&	1.000\\
&&2.0&	0.038&	0.801&	1.000\\
\cline{2-6}
&\multirow{5}{*}{20}&0.0	&0.049&	0.593&	0.918\\
&&0.5&	0.047&	0.599&	0.924\\
&&1.0&	0.046&	0.602&	0.928\\
&&1.5&	0.044&	0.589&	0.924\\
&&2.0&	0.038&	0.545&	0.904\\
\hline
\end{tabular}
\end{subtable}%
\hspace{2.5cm}
    \begin{subtable}[t]{.45\linewidth}
      \centering
        \caption{Uniform dist.}\label{tab:Res2}
\begin{tabular}{ccc}
\multicolumn{3}{c}{$\theta$}\\
0&\multicolumn{2}{c}{1}\\
\hline
&\multicolumn{2}{c}{$p$}\\
\cline{2-3}
&1&2\\
\hline
	0.047&	0.592&	0.998\\ 
	0.044&	0.591&	0.999\\
	0.043&	0.586&	0.999\\
	0.040&	0.595&	0.999\\
	0.034&	0.621&	0.998\\
\hline
	0.047&	0.309&	0.524\\
	0.045&	0.298&	0.522\\
	0.043&	0.285&	0.516\\
	0.040&	0.287&	0.524\\
	0.034&	0.300&	0.545\\
\hline
	0.047&	0.195&	0.208\\
	0.045&	0.184&	0.197\\
	0.043&	0.174&	0.187\\
	0.040&	0.173&	0.187\\
	0.034&	0.176&	0.191\\
\hline
	0.049&	0.992&	1.000\\
	0.047&	0.993&	1.000\\
	0.046&	0.994&	1.000\\
	0.045&	0.996&	1.000\\
	0.042&	0.997&	1.000\\
\hline
	0.049&	0.803&	1.000\\
	0.047&	0.800&	1.000\\
	0.046&	0.800&	1.000\\
	0.045&	0.824&	1.000\\
	0.042&	0.874&	1.000\\
\hline
	0.049&	0.559&	0.901\\
	0.047&	0.546&	0.903\\
	0.046&	0.533&	0.908\\
	0.044&	0.557&	0.924\\
	0.041&	0.627&	0.946\\
\hline
\end{tabular}
\end{subtable}%
\end{table}

\begin{table}[]
\caption{Average proportion of ties per simulation scenario}\label{tab:Resb}
\begin{subtable}{.45\linewidth}
\centering
\caption{Normal distribution}\label{tab:Resb1}
\begin{tabular}{llrccc}
&&&\multicolumn{3}{c}{$\theta$}\\
&&&0&\multicolumn{2}{c}{1}\\
\cline{4-6}
&&&&\multicolumn{2}{c}{$p$}\\
\cline{5-6}
$n$&$\sigma^{\frac{1}{p}}$&$d/\sigma$&&1&2\\
\hline
\multirow{15}{*}{30}&\multirow{5}{*}{10}&0.0&	0.00&	0.00	&0.00 \\
&&0.5	&	0.28&	0.21&	0.10\\
&&1.0&	0.52&	0.4	&0.19 \\
&&1.5&	0.71&	0.57&	0.28\\
&&2.0&	0.84& 0.71&0.36 \\
\cline{2-6}
&\multirow{5}{*}{15}&0.0	&0.00&	0.00	&0.00\\
&&0.5&	0.28	&0.24	&0.17\\
&&1.0&	0.52&	0.46&	0.33\\
&&1.5&	0.71&	0.64&	0.48\\
&&2.0&	0.84&	0.78&	0.61 \\
\cline{2-6}
&\multirow{5}{*}{20}&0.0&		0.00	&0.00&	0.00\\
&&0.5&	0.28&	0.25&	0.23\\
&&1.0&	0.52&	0.48	&0.43\\
&&1.5&	0.71&	0.67	&0.61\\
&&2.0&	0.84&	0.80	&0.75\\
\hline
\multirow{15}{*}{20}&\multirow{5}{*}{10}&0.0&	0.00	&0.00&	0.00\\
&&0.5&	0.28&	0.24	&0.17\\
&&1.0&	0.52	&0.46&	0.33\\
&&1.5&		0.71&	0.64&	0.48\\
&&2.0&		0.84&	0.78	&0.60\\
\cline{2-6}
&\multirow{5}{*}{15}&0.0&	0.00	&0.00	&0.00\\
&&0.5&	0.28	&0.26	&0.24\\
&&1.0&	0.52&	0.49	&0.46\\
&&1.5&	0.71	&0.67&	0.64\\
&&2.0&	0.84&	0.81	&0.78\\
\cline{2-6}
&\multirow{5}{*}{20}&0.0	&	0.00&	0.00	&0.00\\
&&0.5&	0.28&	0.27	&0.26\\
&&1.0&0.52&	0.50&	0.50\\
&&1.5&	0.71&	0.69	&0.69\\
&&2.0	&0.84	&0.82&	0.69\\
\hline
\end{tabular}
\end{subtable}
\hspace{2.5cm}
    \begin{subtable}{.45\linewidth}
      \centering
        \caption{Uniform dist.}\label{tab:Resb2}
\begin{tabular}{ccc}
\multicolumn{3}{c}{$\theta$}\\
0&\multicolumn{2}{c}{1}\\
\hline
&\multicolumn{2}{c}{$p$}\\
\cline{2-3}
&1&2\\
\hline
0.00&	0.00&	0.00\\
0.27&	0.20	&0.10 \\
	0.49	&0.39	&0.19\\
	0.68	&0.56	&0.28\\
	0.82	&0.70&	0.36\\
\hline
	0.00&	0.00	&0.00\\
	0.27&	0.23	&0.17\\
	0.49	&0.44	&0.33\\
	0.68&	0.62&	0.48\\
0.82	&0.76	&0.61\\
\hline
	0.00	&0.00	&0.00\\
0.27	&0.24&	0.22\\
	0.49	&0.46&	0.42\\
	0.68	&0.65	&0.60\\
	0.82	&0.79	&0.74\\
\hline
0.00	&0.00&	0.00\\
	0.27&	0.23&	0.17\\
	0.50	&0.44	&0.33\\
0.68	&0.62&	0.47\\
	0.82&	0.76&	0.60\\
\hline
	0.00&	0.00&	0.00\\
	0.27	&0.24&	0.23\\
	0.49	&0.47	&0.44\\
	0.68&	0.65&	0.62\\
0.82	&0.8	&0.77\\
\hline
	0.00&	0.00&	0.00\\
0.27	&0.25&	0.25\\
	0.49&	0.48&	0.48\\
	0.68	&0.66	&0.66\\
	0.82&	0.81&	0.80\\
\hline
\end{tabular}
\end{subtable}%
\end{table}

\subsection{Choice of value for $d$}\label{Choice of value for $d$}

The value of $d$ should ideally reflect a known data-generating mechanism which involves a known measurement error. However, as noted in Section \ref{sec:dis}, the value of $d$ might substantially impact the sampling distribution of $S_{\rm ex}$ and, as a consequence, $t$. A relatively large $d$ will result in a large number of ties, which in turn might impact the normality of the test statistics and thus invalidate inference. Also, the power of the test will decrease. As shown in Table \ref{tab:Resb1}, the proportion of ties ($\pi_t$) can be substantial, depending on the fraction $d/\sigma$ and the data generating mechanism. We can there see that under $H_0$, $\pi_t$ is independent of sample size and only varies with $d/\sigma$. 

To provide guidance with respect to the choice of the value of $d$ we will further study and disentangle the effect of sample size on the normality of the test statistics. We will employ the level of preservation of the type I error close to the nominal level, as a way of assessing normality. To that end, we conducted an additional simulation study where we assess empirical type I error as a function of the effective sample size ($ESS$) resulting from the application of various levels of $d$. We define $ESS=(1-\pi_t)\times N$, the number of non-tied observations. Under $H_0$, we assess the type I error for $N=\{20,30,50,100\}$ and $d/\sigma \in [0,2]$. We present the results as a function of the $ESS$ resulting from the combination of $N$ and $d/\sigma$.

\begin{figure}[h]
	\centering
		\includegraphics[width=1\textwidth]{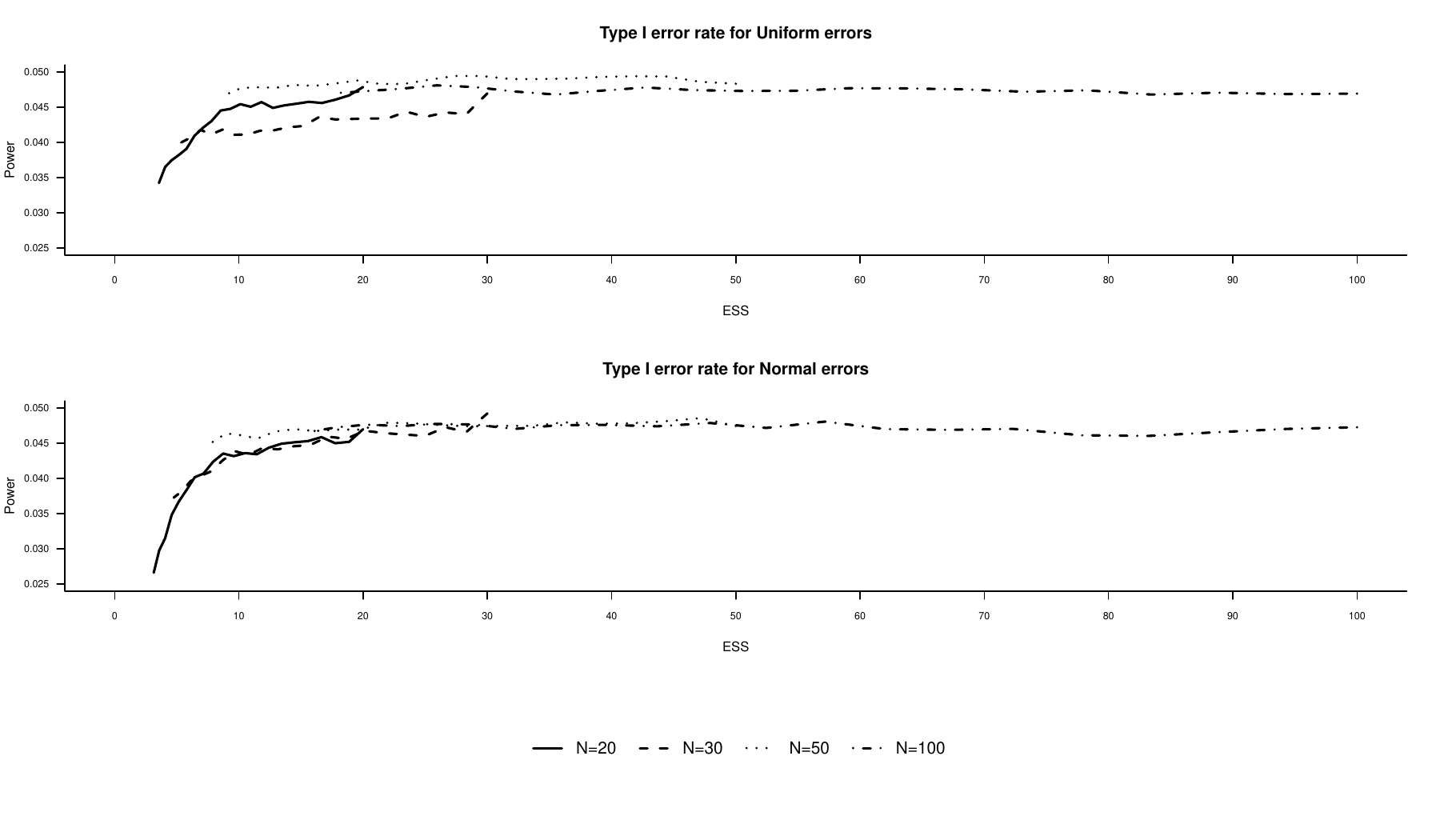}
	\caption{Type I error as a function of $ESS$, for Uniform and Normal errors.}
	\label{fig:plot_ESS}
\end{figure}

Figure \ref{fig:plot_ESS} presents the results of these simulations ($3\times10^4$ runs). It can be found that a $ESS$ of 10 would be a realistic minimum for ensuring that the normal approximation would adequately hold. We can see that this is irrespective of sample size. This also points out why neither $N$ nor $\pi_t$ alone should guide the choice of $d$, as even for a large $\pi_t$ a large enough $ESS$ can be sufficient. 

Making sure that distributional assumptions hold should be a concern when choosing the value of $d$, but not the only one. If measurement error of a known magnitude occurs in the data this should lead the choice for the value of $d$, while the $ESS$ can be monitored at the analysis stage. It should be noted that the effect of $\pi_t$ on the sampling distribution and operational characteristics of the test is expected to be independent of whether $H_0$ holds or not. We use

\section{Application}
We demonstrate the practical use of the extended Mann-Kendall test by applying it to the blood donation data introduced in Section \ref{sec:intro}. We limit the application to the first five years of the study (2001-2005) during which measurement uncertainty was of greater concern \citep{janssen2015}.
 
A subset of the data included in this report (number of donated platelet therapeutic units per 1000 inhabitants) was chosen for demonstrating the extended Mann-Kendall test in an adapted version, namely the Regional Kendall test for trend \citep{regional}. The Regional Kendall test for trend uses test results from individual regions in an overall test to decide upon the existence of a trend across these regions. It does so for combining trends in individual regions as the Seasonal Kendall test for trend does for different seasons \citep{hirsch}. If data are available over $m$ regions for the same time periods, the overall statistic, $S_r$ is computed as follows:

\begin{equation}
S_r=\sum_{L=1}^m S_L,
\end{equation}
where $S_L$ denotes the $S$ statistic calculated from the reported annual data of the $L$'th region. Its variance is the sum of the regional variances and the corrected score, $Z_r$ is computed as above (Equation \ref{eq:Z}). When the product of the number of regions and number of periods is larger than 25 \citep{hirsch}, the distribution of $S_r$ can be approximated by a normal distribution and $Z_r$ can be used for testing against trend.

Table \ref{T3} presents the data, considering countries without missing values only. 

{\begin{table}
\caption{Platelets donated per 1000 inhabitants for countries with data available for all 5 years.}
\centering
\label{T3}
\begin{tabular}{l r r r r r}
\hline
\\
Country & \multicolumn{5}{c}{Year} \\
&2001&2002&2003&2004&2005\\
\cline{2-6}
Belgium& 4.65& 4.57& 4.87& 5.82& 6.98\\
Bulgaria& 1.38& 2.02& 2.16& 0.71& 0.70\\
Czech Republic& 2.11& 2.29& 2.19& 2.37& 2.10\\
Finland& 6.67& 6.66& 6.07& 6.17& 6.42\\
France& 3.26& 3.22& 3.32& 3.35& 3.50\\
Germany& 3.27& 3.27& 4.04& 4.53& 4.45\\
Greece& 11.67& 12.88& 12.63& 15.85& 14.95\\
Iceland& 3.21& 2.42& 3.38& 3.17& 3.66\\
Italy& 8.26& 12.84& 3.37& 2.16& 2.75\\
Latvia& 1.06& 1.70& 1.80& 1.66& 1.74\\
Netherlands& 9.39& 3.45& 2.92& 3.23& 3.19\\
Norway& 3.43& 3.12& 2.99& 3.48& 3.39\\
Poland& 0.94& 1.49& 1.04& 1.30& 1.62\\
Romania& 1.57& 1.84& 1.84& 2.72& 3.13\\
Slovak Republic& 1.36& 1.72& 1.60& 1.60& 1.87\\
Slovenia& 14.00& 11.92& 11.07& 13.08& 13.96\\
Sweden& 3.52& 3.79& 3.63 &3.90& 3.67\\
Switzerland& 2.43& 2.02& 3.12& 2.51& 2.69\\
United Kingdom& 4.50& 4.42& 4.53&4.44& 4.39\\
\hline
\end{tabular}
\end{table} }

The results presented here are intended to illustrate the use of the extended Mann-Kendall test for trend and not to contradict or doubt the results presented in the original report. Let us demonstrate some possible ways of applying the extended Mann-Kendall test and the consequences for conclusions drawn.

First, a simple Regional Kendall test (so $LRD$=0) would yield a value for $S_r$ of 41 and a two-sided $p$-value of approximately 0.024. Testing on the 0.05 significance level, one would conclude that there is a significant upward trend in platelets donated within the European Union in the period of 2001-2005. When one would consider a difference of 0.4 platelets donation per 1000 inhabitants not meaningful, the $p$-value would be 0.07 and one would conclude that the presence of a trend would not be statistically significant. Alternatively, one could also argue that a different $LRD$ per country would be more appropriate. For example, it would make sense to apply a different $LRD$ to the data from Greece, where on average there are 13.6 platelets transfused per 1000 inhabitants, than to the data from Bulgaria, where on average there are 1.39 platelets transfused per 1000 inhabitants. So, an $LRD$ as a percentage of the average number of platelets transfused per 1000 inhabitants could be chosen as an alternative. The results of the Regional Kendall test for trend when a 0.05, 0.10 and 0.20 of the average reported number of platelets transfused per 1000 inhabitants is used as an $LRD$ alongside with the results discussed above are presented in Table \ref{T4}.

We see that the choice of $LRD$ is crucial for the results obtained. A different $LRD$ will lead to different results as far as the significance of the trend is concerned. Therefore it is recommended that prior knowledge on uncertainty in the observed values or consensus on a meaningful difference in subsequent observations is obtained before applying the extended Mann-Kendall test.

{\begin{table}
\caption{Results of Regional Kendall test applying extended Mann-Kendall test per country. $LRD$=0 corresponds to the standard Mann-Kendall test.}
\centering
\label{T4}
\begin{tabular}{l  c c c}
\hline
\\
LRD & \multicolumn{3}{c}{Results} \\
&$S_r$&Var$(S_r)$& 2-sided\\
&&&$p$-value\\
\cline{2-4}
0& 41& 315.67& 0.0244\\
0.05& 45& 295.67& 0.0105\\
0.20& 41& 223.67& 0.0075\\
%0.40& 24& 163.33& 0.0719\\
5\% of c.m.$^*$ &49& 239.67& 0.0019\\
10\% of c.m.& 41& 175& 0.0025 \\
%20\% of c.m.& 21& 113.67& 0.0607\\
\hline
{\scriptsize{$^*$countries' mean}}\\
{\scriptsize{platelet use}}
\end{tabular}
\end{table} }

\section{Discussion}

In this paper, we propose the use of an extension of the Mann-Kendall test for trend to allow for measurement error and partial ties. That is, we propose that in the calculation of the Mann-Kendall test statistic any values between which the difference is considered not to be meaningful are interpreted as ties. In practice an $LRD$ is applied such that in case the difference between two values is less than some value $d$ these values are considered equal. This allows the Mann-Kendall test to account for measurement error and/or meaningful differences between observed outcomes. The idea of partial ties could essentially be extended to any rank-based test, even though this is beyond the scope of the present manuscript.

We derived a formula for the variance of the extended statistic and provide analytical results for a class of contiguous alternatives. In addition, we demonstrated the behavior of the test in the presence of a linear or quadratic trend with normally or uniformly distributed errors via a simulation study. The operational characteristics of the extended Mann-Kendall test are appealing compared to the conventional Mann-Kendall test as the extended test increases power while retaining the type I error to nominal levels.

We demonstrated that for small $d$ there is (asymptotically) a marginal gain in power for the class of contiguous alternatives studied. This gain is shown to be a function of the error variance when the errors are normally distributed. Via simulation we explored linear and quadratic alternatives for normal and uniform errors, and showed that there can be a substantial increase in power in case distributed errors follow a uniform distribution. Our main conclusion is that the extended test will be useful once a user has a good understanding of the data generating mechanism and measurement error. 

Attention should be paid to the number of ties when applying an $LRD$. Applying a relatively large $LRD$ can lead to a substantial number of tied values in the data. In extreme cases this may lead to implications concerning the normal approximation of the distribution of $S_{\rm ex}$. In such cases a permutation test will be more appropriate in order to acquire exact $p$-values which would always be valid for such a model \citep{ernst2004}. Through simulation we demonstrated that the normal approximation can be safely assumed to hold if the resulting $ESS$ is larger than 10. Even though we assessed these properties under $H_0$, since nominal power levels are known and we can use simulations to assess the tests' performance, it should be noted that the effect of $\pi_t$ on the sampling distribution and operational characteristics of the test is expected to be independent of whether $H_0$ holds or not. Thus the guidance for $ESS >$ 10 is given independent of the truth of $H_0$.

As shown in the examples presented, the proposed method approach can also be used when combining multiple Mann-Kendall tests (Seasonal Kendall and the Regional Kendall test for trend). In this case extra attention should be paid so the $LRD$ applied is realistic for a specific dataset. Different conclusions can result from different values of $LRD$ and therefore $d$ should be meaningful and case-specific.

It should be noted that the focus of this work is on hypothesis testing and we only worked out the distribution of $S_{ex}$ and not of the resulting $t$. Thus, issues related to point or interval estimation of $\tau$ (bias, confidence intervals or bands) call for further research. 

The extension can be useful for applied researchers in many disciplines were non-parametric testing for trend is applied. It requires knowledge of the measurement error in the observed data, but will contribute to more robust and meaningful conclusions regarding the existence of a trend over time. R code for calculating the extended version of the test is publicly available on the first author's github repository (https://github.com/SNikolakopoulos/ExtendMK).

\section*{Acknowledgements}
The paper is published at \textit{
Statistics: A Journal of Theoretical and Applied Statistics, doi 10.1080/02331888.2023.2214942}. 	We would like to thank the Associate Editor and two anonymous reviewers. Their insightful comments have substantially improved the presentation of the work.

%%%%%%REFERENCES%%%%%%%
\bibliographystyle{chicago}
\bibliography{MKext}

\end{document}